\documentstyle[twocolumn,prc,aps,epsfig]{revtex}
\input epsf
\newcommand{\be}{\begin{eqnarray}}
\newcommand{\ee}{\end{eqnarray}}

\begin{document}

\twocolumn[\hsize\textwidth\columnwidth\hsize\csname @twocolumnfalse\endcsname

\title{Spin-Spin and Spin-Orbit Interactions\\
in Strongly Coupled Gauge Theories } 

\author {Edward V. Shuryak and Ismail Zahed}
\address { Department of Physics and Astronomy\\ State University of New York,
     Stony Brook, NY 11794-3800}

\date{\today}
\maketitle
\begin{abstract}
We evaluate the spin-orbit and spin-spin interaction between two
fermions in strongly coupled gauge theories in
their Coulomb phase. We 
use the quasi-instantaneous character of Coulomb's law at strong
coupling to resum a class of ladder diagrams. For ${\cal N}=4$ SYM 
we derive both weak and strong coupling limits of the
  the spin-orbit and spin-spin interactions,
and find that in the latter case these interactions
 are subleading corrections and do not seriously affect
 the deeply bound Coulomb states with large angular momentum,
 pointed out in our previous paper.
The results are important
for understanding of the regime of intermediate coupling, which is the case
for
 QCD somewhat above the chiral transition temperature. 
\end{abstract}
\vspace{0.1in}
]
\begin{narrowtext}
\newpage


{\bf Introduction.\,\,}
QCD at temperatures $T=(1-3)T_c$ is in a chirally restored but 
relatively strongly coupled
 Coulomb phase. Although it is known for long time that some aspects of
this regime are nonperturbative ( see \cite{Shu_80,nonp}
and references therein), and 
although explicit evaluation of perturbative series for
thermodynamical quantities are not convergent in this domain,
till recent times one have mostly relied on perturbative ideas
for signal assessment 
at RHIC. Recently, we have suggested~\cite{sz03} that this region
of temperatures 
supports $\bar q q$ (light and heavy) and $gg$ 
Coulomb bound states, enhanced due to running of the coupling to
smaller momentum scale as compared to the vacuum.
 Charmed and even light flavor mesonic
states beyond the chiral restoration
transition have been also observed on the lattice. 
In \cite{sz03} we further suggested that the
 scattering cross section peak near  the lines of zero
binding energy, and reach  large values, explaining why
the QGP at RHIC behaves hydrodynamically.

The nonperturbative character of the Coulomb phase
can be further investigated in ${\cal N}=4$ SYM in the
Maldacena limit~\cite{maldacena}, at parametrically strong coupling
 $\lambda=g^2N \gg 1$ (where $N$ is the
 number of colors)
and finite temperature $T$. In particular, the potential between 
two static Coulomb charges is found to obey a modified  Coulomb's 
law~\cite{maldacena,rey}. ${\cal N}=4$ SYM at finite temperature and
in strong coupling is similar to QCD across the chiral restoring
temperature $T_c$. We recall, that a number of pertinent issues have
been addressed in strongly coupled ${\cal N}=4$ SYM, such as
small angle scattering~\cite{zahed,janik},
large angle scattering~\cite{polchinski}, the free
energy~\cite{thermo}, the electric Debye screening~\cite{Rey_etal},
the viscosity~\cite{PSS}, and real-time correlators~\cite{Starinets}.

The puzzling aspect of all the finite temperature results 
is that they are independent
of the strong coupling constant $\lambda=g^2N_c$, even though the 
(modified) Coulomb interaction is proportional to $\sqrt{\lambda}$.
The naive quasiparticles carry thermal Coulomb energies
of order $\sqrt{\lambda}\,T$, making them way too heavy to be thermally
excited  in the Coulomb phase. Hence, what is the finite-T
matter in the strongly coupled Coulomb phase made of?
The answer to this question was suggested in \cite{sz03}, where it was
pointed out that in the strong coupling regime, there exists about a
constant density of deeply bound
composite states of mass $m\approx T$ at
large $\lambda$. These composites are made of two quasiparticles which
rotate with a large angular momentum $l\approx \sqrt{\lambda}$,
compensating the supercritical Coulomb force.

The present paper addresses the following issue.
In~\cite{sz03} the spectrum of the composites was limited to the 
simplest case of spinless scalars obeying the Klein-Gordon equation.
The cases of composites made of two fermions and/or gluons were only
discussed qualitatively. One may wander what exactly is the role of
the spin-spin and spin-orbit interactions in these highly relativistic
states, especially since the orbital momentum is parametrically large.
In this letter we answer this question by extending 
the arguments of \cite{sz03} to the spin-spin and spin-orbit
interactions at strong coupling: we well show that these forces
are subleading in the strong coulping limit.

In principle, this question can be exactly addressed using the AdS/CFT
correspondence by evaluating the averages of pertinent supersymmetric
Wilson lines for spinning and moving (non-static) particles along
externally chosen paths. Such approach should yield exact answers with
exact coefficients. Alternativaly, one can also obtain parametrically
correct answers (without exact coefficients) using the resummation of
ladder diagrams. This was first demonstrated for the modified
Coulomb's law in~\cite{zarembo}, see also the discussion of the higher
order and screening effects in~\cite{sz03}. The key physics idea is that 
in the strong coupling regime gauge (scalar) exchanges between two
charges separated by a distance $L$ takes place during a short time
$L/\lambda^{1/4}$, in contrast to the weak coupling regime where it
is of order $L$.

After briefly recalling the derivation of the modified Coulomb's law
in ${\cal N}=4$ SYM using the ladder resummation, we introduce the
correlator of two  ``spinning''
fermionic lines. In the relativistic case
 the mass in the magnetic moment is substituted by the (relativistic)
kinetic energy of the external particle on its classical orbit. We use
a ladder resummation both in weak and strong coupling to derive the spin
induced interactions between spin $1/2$ particles. In weak coupling, we
reproduce the familiar Breit-Fermi interaction, while in strong coupling
a novel spin-induced interaction is derived. The spin-induced
interactions are used to qualitatively assess the spin splittings in QCD 
across the chiral restoring temperature.

\vskip 0.5cm


\begin{figure}[h!]
 \centering
 \includegraphics[width=6cm]{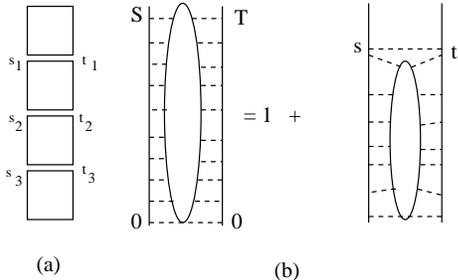}
 \caption{\label{fig_ladders} 
 (a) The color structure of ladder diagrams in the large-$N_c$
 limit: each square is a different color trace, bringing the factor
 $N_c$.
  The time goes vertically, and the planarity condition
  enforces strict time ordering, $s_1>s_2>s_3...$,
 $t_1>t_2>t_3...$. (b) Schematic representation of the Bethe-Salpeter 
 equation (\ref{eqn_BS}) summing ladders.
 }
 \end{figure}

{\bf The modified Coulomb's law.\,\,}
In the limit where the number of colors $N_c$ is large and
the coupling $\lambda=g^2N_c$ is large as well, charges in
${\cal N}=4$ SYM communicate with quasi-instantaneous and 
nearly abelian interactions. This legitimates the use of
ladder diagrams to understand the nature of the strong
interaction between heavy~\cite{zarembo} and light charges.

The quasi-instantaneous potential between two charges at a 
distance $L$, can be obtained by resumming the ladder diagrams
 displayed in Fig. \ref{fig_ladders}.  
The pertinent Bethe-Salpeter kernel is 
solution to the integral equation

\be \label{eqn_BS}
\Gamma({\cal S},{\cal T})=1+{\lambda \over 4\pi^2}\int_0^{\cal S} ds
\int_0^{\cal T} dt{1\over
    (s-t)^2+L^2} \Gamma(s,t)\,\,.
 \label{1a}
\ee
The first factor under the integral is the (Euclidean) propagator for
one extra gluon/scalar added to the ladder. The kernel 
satisfies the boundary condition $\Gamma ({\cal S},0) =\Gamma(0,{\cal T})=1$.
when the equation is solved, the ladder-generated potential follows from

\be
V_{\rm lad}(L) 
=-\lim_{T\to{+\infty}}{\frac 1{\cal T} \Gamma\, ({\cal T},{\cal T})}\,\,,
\label{0a}
\ee

In weak coupling $\Gamma\approx 1$  and the integral on the rhs is
easily taken, resulting in 

\be
\Gamma ({\cal S,T}) \approx 1+\frac{\lambda}{8\pi}\,\frac{{\cal S+T}}L
\label{00a}
\ee
which results into the standard Coulomb's law.
Note that in this case the typical relative time difference
between emission and absorption of a quantum $|t-s|\approx L$, so one can
say that virtual quanta travel at a speed $v\approx 1$. In strong
coupling, the kernel can be obtained using the method outlined in
\cite{zarembo}. For $x={\cal (S-T)}/L$ and $y={\cal (S+T)}/L$, the
result for large times ${\cal S\approx T}$ that is small $x$ and large 
$y$ is

\be
\Gamma (x,y)\approx {\bf C}_0\,e^{-\sqrt{\lambda}\,x^2/4\pi}\,
e^{\sqrt{\lambda}\,y/2\pi}\,\,.
\label{5a}
\ee
From (\ref{0a}) it follows that
in the strong coupling limit the ladder generated potential
is 

\be V_{\rm lad}(L)= -\frac{\sqrt{\lambda}/\pi}L \ee which 
has the same parametric form  as the one derived from the
AdS/CFT correspondence \cite{maldacena,rey}  except for the
overall coefficient. The discrepancy is due to the left out
higher-order diagrams discussed in~\cite{sz03}.

For physics it is important to stress again that in the strong coupling
limit $\lambda=g^2N\gg 1$, the color charges communicate on
short time of order $L/\lambda^{1/4}$ within which the exchange
is nearly Abelian. This allows the use of a potential description {\it
even} for relativistically moving charges~\cite{sz03}. So the two
particle relativistic problem is emmenable to a bound state problem
with a potential at strong coupling.

\vskip 0.5cm


{\bf Non-static Wilson lines with spin.\,\,}
To investigate the spin effects on the form of the potential
inherited by the resummation of the ladders, we note that the 
basic correlator between say two relatistic spin $\frac 12$ 
particles follows from the generic correlation function

\be
&&{\bf C} ({\cal T}) =\nonumber\\
&&
\Bigl<{\rm Tr}\,{\bf P}\,{\rm exp}\left(g\int_0^{\cal T}\,ds_1\,(i\,\dot{x}_1\cdot A (x_1)
+\frac 1{4} \sigma_{1\mu\nu}\,F_{\mu\nu} (x_1))\right)\nonumber\\
&&\times
{\rm exp}\left(-g\int_0^{\cal T}\,ds_2\,(i\,\dot{x}_2\cdot A (x_2)
+\frac 1{4} \sigma_{2\mu\nu}\,F_{\mu\nu} (x_2))\right)\Bigr>\,\,,
\label{cor}
\ee
with {\it fixed} external paths. The new spin-dependent additions are the
$\sigma_{1,2}$ terms acting on the spinor indices of both fermionic
lines. The path ordering ${\bf P}$ and the trace is over spin and color.
The quantum propagator follows from the integration over the external
paths $x_{1,2}$ and the affine time ${\cal T}$.
Gauge invariance is enforced for trajectories with identical 
end-points $x_1=x_2$ at $s=0,{\cal T}$. The averaging 
is over the pertinent gauge field measure.

Supersymmetric spinning lines can be also discussed, but we will not
do it here. Since our spinning lines are just {\rm external probes}
rather than part of the underlying gauge theory, their color charges
can be chosen at will, e.g. in the fundamental representation. We
will also ignore the scalar exchanges, so the correlator is not even
supersymmetric. Thus it has divergent self-energy corrections, which
however drop out of the interaction potential which we will be 
calculating below. We further note that the two spinors are chosen
to move along the {\it fixed} external paths, which we can select at will
to get the potential we need. The quantum correlator for spin $1/2$ 
follows from (\ref{cor}) by integrating over the external paths 
$x_{1,2}$, which will not be discussed below.

The (diffeomorfic) time integration in (\ref{cor})
is over the affine time $s$ which has dimensions of
{\rm inverse mass square}. If one uses the 
laboratory time $t$ in (\ref{cor}), 
 a relativistic gamma factor of the particle (due to Lorentz
invariance) appears, and so
  $S=m\int dt \sqrt{1-\dot{x}^2}$ (in Minkowski notations). 
Starting with the action of a free particle with affine
time $s$ 

\be
{\bf S} =\int_0^{\cal T}\,ds \,\frac 12 (\dot{x}^2+m^2)
\label{free}
\ee
one may rewrite it in terms of the conventional time $t$ 
using additional parameter $\mu=dt/ds$
\be
{\bf S} =\int_0^{T}\,dt \,\left(
\,\frac \mu{2}\,(1+\dot{\vec x}^2)+\frac {m^2}{2\mu}\right)
\label{free1}
\ee
Note that $\mu$ in general depends
on the particular trajectory. For a free particle it can be set
by extremizing the action over $\mu$ (\ref{free1}) \cite{simonov},
with the result 
\be
\mu =\frac{m}{\sqrt{1+\dot{\vec x}^2}} =\gamma\, m\,\,,
\label{free2}
\ee
which is just the relatistic energy of a free particle,
since the relativistic momentum in Euclidean space is
$\vec{p}=i\gamma\,m\dot{\vec{x}}$.

In terms of the physical
time, the correlation function (\ref{cor}) now reads

\be
&&{\bf C} ({T}) =\nonumber\\
&&
\Bigl<{\rm Tr}\,{\bf P}\,{\rm exp}\left(g\int_0^{T}\,dt_1\,(i\,\dot{x}_1\cdot A (x_1)
+\frac 1{4\mu} \sigma_{1\mu\nu}\,F_{\mu\nu} (x_1))\right)\nonumber\\
&&\times
{\rm exp}\left(-g\int_0^{T}\,dt_2\,(i\,\dot{x}_2\cdot A (x_2)
+\frac 1{4\mu} \sigma_{2\mu\nu}\,F_{\mu\nu} (x_2))\right)\Bigr>\,\,.
\nonumber\\
\label{corx}
\ee
In the chiral basis, the spin term reads

\be
\frac g{4\mu} \,\sigma_{\mu\nu}\,F_{\mu\nu}
=\frac g{2\mu}\vec{\sigma}\cdot\left(\vec{B}\,{\bf 1_+}-\vec{E}\,{\bf 1_-}\right)
\label{sf}
\ee
with ${\bf 1_\pm}={\rm diag}(1,\pm 1)$. We note that the positive
parity of the electric field is flipped by the chirality matrix
${\bf 1_-}$. The spin-coupling to the magnetic field occurs with
the correct magnetic moment. The spin-coupling to the electric field 
is purely imaginary for physical electric fields\footnote{Note
that it is 
real for virtual vacuum fields such as instantons.}. This point is
particularly important in deriving the spin induced interactions,
since the electric parts generate mostly phases that drop from the
positive correlation function (\ref{corx}).

In strong coupling the ladder approximation maybe used to assess the
nature of the spin forces. For that, the driving kernel is given by
the {\it real parts} of the expectation value

\be
&&\Bigl<\left(g\,(i\,\dot{x}_1\cdot A (x_1)
+\frac 1{2\mu} \vec{\sigma}_1\cdot \vec{B} (x_1))\right)\nonumber\\
&&\times
\left(-g\,(i\,\dot{x}_2\cdot A (x_2)
+\frac 1{2\mu} \vec{\sigma}_2\cdot \vec{B}  (x_2))\right)\Bigr>\,\,,
\label{ex}
\ee
where the imaginary contributions through the electric field have been
dropped.
Again the Coulomb kernel arises 
from the $AA$ correlator in the form

\be
\frac {g^2\,C_A}{4\pi^2}\,\frac{1+\dot{\vec{x}_1}\cdot\dot{\vec{x}_2}}
{(t_1-t_2)^2+(\vec{x}_1-\vec{x}_2)^2}
\label{AA}
\ee
with the Casimir $C_A=N/2$ in large $N$. Note that the velocity
dependent part translates in Minkowski space to $(1-\vec{v}_1\cdot
\vec{v}_2)$ which is the expected correction to Coulomb's law from
Ampere's law as induced by charge motion. This contribution adds
for particle-antiparticle or particle-hole, and subtracts otherwise
in relativistic bound states. For instance, for relativistic
particle-antiparticle states  $v_1=-v_2$, twicing the Coulomb 
interaction. At asymptotic temperature Ampere's induced interaction 
can still be used space-like to bind in the dimensionally reduced
theory~\cite{detar}.

\vskip .5cm

{\bf Spin-dependent contributions.\,\,\,}
The spin-orbit contribution follows from the $AB$ term in the form

\be
\Bigl<\left(ig\,\dot{\vec{x}}_1\cdot\vec{A}(x_1)\right)
\left(\frac {-g}{2\mu} \vec{\sigma}_2\cdot \vec{B}  (x_2))\right)\Bigr>
+1\leftrightarrow 2
\label{AB1}
\ee
Thus

\be
-\frac{g^2\,C_A}{4\mu^2}
\left(\vec{\sigma}_2\cdot(\vec{x}_{12}\times \vec{p}_1) 
+\vec{\sigma}_1\cdot(\vec{x}_{21}\times \vec{p}_2) \right)
\frac 1{\pi^2\,x^4}
\label{AB2}
\ee
where again we have used 
$\vec{p}=i\gamma\,m\dot{\vec{x}}$.

The spin-spin contribution follows from the $BB$ term in the form

\be
-\frac{g^2}{4\mu^2}
\sigma_{1i}\sigma_{2j}\,
\Bigl<B_i(x_1)\,B_j(x_2)\Bigr>\,\,.
\label{AB3}
\ee
The result is

\be
\frac {g^2\,C_A}{4\mu^2}
\left(\vec{\sigma}_1\cdot\vec{\sigma}_2\,\nabla^2-
\vec{\sigma}_1\cdot\vec{\nabla}\,
\vec{\sigma}_2\cdot\vec{\nabla}\,\right)
\,\frac 1{4\pi^2\,x^2}\,\,.
\label{AB4}
\ee

The spinning ladder diagrams can be generated in the same way
as the non-spinning ones described above. In Feynman gauge, the 
driving kernel in (\ref{1a}) is now

\be
&&+\frac{\lambda}{8\pi^2\,x^2} \left( 1-\frac
{\vec{p}_1\cdot\vec{p}_2}{\mu^2} 
\right)\nonumber\\
&&+\frac{\lambda}{8\pi^2\,x^4}\frac 1{\mu^2}
\left(\vec{\sigma}_2\cdot(\vec{x}_{21}\times \vec{p}_1) 
+ 1\leftrightarrow 2\right)\nonumber\\
&&+\frac{\lambda}{32\pi^2\,\mu^2} 
\left(\vec{\sigma}_1\cdot\vec{\sigma}_2\,\nabla^2-
\vec{\sigma}_1\cdot\vec{\nabla}\,
\vec{\sigma}_2\cdot\vec{\nabla}\,\right)\,\frac 1{x^2}
\label{2}
\ee
Note that the spin-independent Coulomb term is $1/2$ its
value in (\ref{1a}) since there is no scalar exchange
between the non-supersymmetric external lines in (\ref{cor}).

\vskip .5cm

{\bf Spin-orbit from Thomas precession.\,\,\,}
The spin-orbit term induced by Thomas precession is not present
in our analysis of (\ref{cor}). For non-accelerating paths, this effect
is not even visible. To see it, we note that the spin precess along the 
trajectory with a rate given by the spin factor~\cite{strominger}

\be
{\bf P}\,{\rm exp}\left(-i\int_0^{\cal T}\,d\tau\,\frac 1{2}
\sigma_{\mu\nu}\dot{x}_\mu\dot{\dot{x}}_\nu\right)\,\,,
\label{fa}
\ee
with $\tau$ the affine proper time with dimensions one over mass.
For a non-accelerating external trajectory, this effect is not
visible. For a particle in an external field, the 4-acceleration
is given by the Lorentz force in proper time

\be
m\dot{\dot{x}}_\nu= g\,F_{\nu\sigma}\,\dot{x}_\sigma\,\,.
\label{lo}
\ee
Inserting (\ref{lo}) in (\ref{fa}) and using 
$\left<\vec{E}\right>=-i\vec{\nabla}\,V_E$ for 
two static charges, it follows that (\ref{fa}) generates

\be
-\frac {g^2\,C_A}{4\mu^2}\,\left(
(\vec{\sigma}_1\times \vec{p}_1\right)\cdot\vec{\nabla}\,V_E({\bf x})
+ 1\leftrightarrow 2 
\ee
which is the standard Thomas contribution (after the
$1/2$ correction in the energy due to the accelerating frame).
This result holds for both weak and strong coupling, with $V_E({\bf x})$ the
pertinent Coulomb potential in strong coupling.

\vskip 0.5cm


{\bf Potentials in weak coupling.\,\,\,}
In weak coupling $\lambda\ll 1$, and the
standard form of the Coulomb and spin interactions can be
obtained by setting $x^2=t^2+\vec{\bf x}_{12}^2$ and integrating
over the time t. The answer is readily found to be

\be
&&{\bf V}_{12} ({\bf x}) =
-\frac{\lambda}{8\pi\,|{\bf x}|} \left( 1-\frac
{\vec{p}_1\cdot\vec{p}_2}{\mu^2} \right)\nonumber\\
&&-\frac{\lambda}{16\pi\,|{\bf x}|^3}\,\frac 1{\mu^2}
\left(\vec{\sigma}_2\cdot(\vec{x}_{21}\times \vec{p}_1)
+1\leftrightarrow 2\right)\nonumber\\
&&+\frac{\lambda}{32\pi\,|{\bf x}|^3} 
\frac 1{\mu^2}
\left(\frac {2|{\bf x}|^3}3\vec{\sigma}_1\cdot\vec{\sigma}_2
\,4\pi\,\delta\,({\bf x})+3\,\sigma_{T12}\right)
\label{3}
\ee
where we have defined the tensor interaction

\be
\sigma_{T12}=\vec{\sigma}_1\cdot\hat{\bf{x}}_{12}
\,\vec{\sigma}_2\cdot\hat{\bf{x}}_{12} -\frac 13 
\vec{\sigma}_1\cdot\vec{\sigma}_2\,\,.
\nonumber
\ee
The presence of $\mu$ makes the potential non-local
in the relativistic limit, since $\mu$ is determined by the
Coulomb trajectory.

The result (\ref{3}) is the standard Breit-Fermi interaction,
modulo the Thomas term as discussed above. For S-states, the
induced interaction is

\be
{\bf V}^{L=0}_{{\rm ladd}\,12} ({\bf x}) =
&&-\frac{\lambda}{8\pi\,|{\bf x}|} \left( 1-\frac
{\vec{p}_1\cdot\vec{p}_2}{\mu^2} \right)\nonumber\\
&&+\frac{\lambda}{3\mu^2} 
\,\vec{S}_1\cdot\vec{S}_2
\,\,\delta\,({\bf x})\,\,.
\label{3x}
\ee
At the critical coupling for S-states $\lambda_c=\pi^2$, the
relativistic Coulomb bound states have a size of order 
$1/\sqrt{\lambda_c}T_c$ for $\mu\approx \pi T_c$~\cite{sz03}, 
smaller than the electric and magnetic screening lengths
(see below and also \cite{sz03}). 
The spin-spin term is then of order

\be
\frac {\pi^2\,T_c}{3}\,\vec{S_1}\cdot\vec{S}_2\,\,.
\label{SS1}
\ee
This causes a splitting between the spin-1 and spin-0
of about $\pi^2\,T_c/3\approx \pi\,T_c$.
However, since the coupling
is rather large, the issue is calling for a strong coupling reassessment 
of the spin-induced interactions.

\vskip 0.5cm


{\bf Potentials in strong coupling.\,\,}
In strong coupling a resummation of the ladders is warranted
as described above. The kernel can be expanded in powers of
relative time $S-T$ to second order, yilding the oscillator-type
Schreodinger equation. The result is the following potential 
\be
&&{\bf V}_{{\rm ladd}\,12} ({\bf x}) =
-\frac{\sqrt{\lambda}}{\pi\,|{\bf x}|} 
\Bigl[ \left(1-\frac{\vec{p}_1\cdot\vec{p}_2}{\mu^2}\right)
\nonumber\\
&&+\frac 1{\mu^2\,|{\bf x}|^2}
\left(\vec{\sigma}_2\cdot(\vec{x}_{21}\times \vec{p}_1) 
+ 1\leftrightarrow 2\right)\nonumber\\
&&+\frac 1{\mu^2\,|{\bf x}|^2}
\left(\frac 13\vec{\sigma}_1\cdot\vec{\sigma}_2-2\,\sigma_{T12}\right)
\Bigr]^{1/2}\,\,.
\label{4}
\ee
where the Coulomb, Amper, spin-spin and spin-orbit terms  all enter
together into the common frequency of the effective oscillator.

{\bf Corrections to Coulomb bound states} due to spin-spin and
spin-orbit terms discussed above are easy to estimate only for 
weakly bound states, for which the non-relativistic limit
$\mu\rightarrow m$ can be used.
In general, the quantity   $\mu=m\gamma$, or the ``
relativistic kinetic energy''
of the particle as we will refer to it,
 is only defined precisely for a given path.

A particular
quantum state can be viewed as a sum of many paths, so one should
average the potential 
over them. 
It is in general a nontrivial problem we are not trying
to solve in this work.
  We will however provide a semiclassical estimate, substituting the
kinetic energy $\mu$ by the total energy minus the potential $E-V$.
Semiclassically, the total energy $E$ is the sum of kinetic ones plus
a potential
\be E=\sqrt{p_1^2+m^2}+ \sqrt{p_2^2+m^2}+V=2\mu+V\ee
and thus the variations of $\mu$ are opposite to those of $V$
as the particle mover from one position to the next.

This clarifies the follwing paradox.
Naively, if one would think of $\mu$ as a constant,
the factors of $1/x \mu$ in spin-spin and spin-orbit terms appear to
signal their dominance at small distances. 
This is not so if it is understood as $\mu \rightarrow E-V$, which 
at
small distances is dominated by the Coulomb potential. Therefore, in
the strong coupling,  
one finds that at small distances it is in fact suppressed
 $1/(\mu\,x)\approx 1/\sqrt{\lambda}\ll 1$. 

 In strong coupling the angular
momentum is of order $\sqrt{\lambda}$. Thus, the spin-orbit coupling is 
of order $1/\sqrt{\lambda}$ while the spin-spin coupling is of order
 $1/\lambda$, both of which are subleading in the strong coupling
limit. 

This justifies {\it a posteriori} the use of the  Coulomb orbit
in the determination of $\mu$ and shows that the results of \cite{sz03}
are left unaffected by spin effects.

 As an example, let us consider the case of
 ultrarelativistic strongly coupled S-states $|p_1|=|p_2|=\mu$. 
The interaction in this case reads
 is
\be
{\bf V}^{L=0}_{{\rm ladd}\,12} ({\bf x}) =
-\frac{\sqrt{\lambda}}{\pi\,|{\bf x}|}
\Bigl[ 1+1 +\frac{4\pi^2}{3\lambda}
\vec{S}_1\cdot\vec{S}_2\Bigr]^{1/2}\,\,,
\label{spin0}
\ee
 To leading order in strong coupling, the 
spin-0 and spin-1 S-states are actually degenerate.
Note also, that the sign of the spin-spin term
is now opposite to the Breit-Fermi one. 

\vskip 0.5cm

{\bf Effects of screening at finite-temperature.\,\,\,}
The above results apply to an unscreened
interaction. In the presence of screening the Coulomb and spin induced
interactions are modified since the electric and magnetic interactiona
are screened as discussed in~\cite{sz03}. In the ladder approximation
and weak coupling, the electric effects are Debye screened at distances
of the order of the inverse Debye lenght $1/m_D\approx
1/\sqrt{\lambda}T$, while the magnetic effects are screened at 
distances of the order of the inverse magnetic lenght $1/m_M\approx
1/\lambda T$. The magnetic length is much larger than the
electric length in weak coupling, 
causing the magnetic effects to be dominant at
asymptotically large temperatures as in QCD.
As a result, magnetically bound Coulomb states may form, see e.g.
 discussion of the so called ``screened masses'' in~\cite{detar}
and more recent discussion of Cooper pair formation
and color supercondctivity \cite{son}.
In strong coupling and in the ladder approximation,
the magnetism is there together with the Coulomb term
in the form of Ampere's term, while the spin-induced effects are
kinematically
suppressed as shown above. String
arguments suggest that the electric and magnetic screening are
comparable in strong coupling, and of order $1/\pi T$ \cite{sz03}.
We have used the latter point of view at $T_c$ in QCD. This point
deserves further investigation.

\vskip 0.5cm

{\bf Summary and discussion.\,\,\,}
 Using the ladder resummation in strong coupling,
we have assessed the spin-spin and spin-orbit potentials  between
two external ``spinning'' fermion paths. The results
in matter follow readily from the screened potentials discussed in
\cite{sz03}. These results can be checked using the ADS/CFT
correspondence from the large time asymptotics of the correlator 
(\ref{cor}) in strong coupling and large $N$. 

In strongly coupled 
${\cal N}=4$ SYM at finite temperature, there are deeply bound Coulomb
 states with orbital angular momentum $\sqrt{\lambda}$, with the total
energy $E\sim T$ and independent on
 $\lambda$, with sizes of the order $1/\sqrt{\lambda}T$. 
These states are bound by ordinary Coulomb (plus Ampere's) law: we have
now shown
that for them the spin-orbit and spin-spin interactions are subleading
effects. 

 These findings are important for evaluation
of related effects in QCD at $T=(1-3) T_c$ in which it
 undergoes a transition into an
 intermediatly coupled Coulomb phase. The pertinent
splitting is the $\rho-\pi$ meson splitting accross the region,
a very important issue for a number of signals at RHIC,
such as the dilepton emission \cite{Shu_80}.

As we shown above, if one uses the weak coupling result --
 the standard Breit-Fermi interaction --
 at the critical
(for the S-states) coupling $\lambda_c=\pi^2$,
it yields a large splitting between the 
spin-0 and spin-1 states of order $\pi\, T_c$.
However our arguments above imply a kinematical suppression by
relativistic factors, on top of which is
 the flipped sign of the spin-spin coefficient,
from weak to strong coupling. It all shows that the issue should
need a special careful examination, both by analytic means and on the lattice.

\vskip 1.25cm
{\bf Acknowledgments}
\\\\
We thank Gerry Brown for discussions.
This work was supported in parts by the US-DOE grant
DE-FG-88ER40388.

\end{narrowtext}

\begin{thebibliography}{99}

\bibitem{Shu_80} E. V. Shuryak, {\it Phys.Rep.} 61 (1980) 71
\bibitem{nonp}
I. Zahed, `Light relativistic bound states in high temperature
QCD', Ed. H. Ezawa, T. Arimitsu and Y. Hashimoto, (1991) North
Holland, p. 357, and references therein.


\bibitem{sz03} 
E.~V.~Shuryak and I.~Zahed, {\tt hep-ph/0307267};
{\tt hep-th/0308073}.

\bibitem{maldacena}
J. Maldacena, Adv. Theor. Math. Phys. {\bf 2} (1998) 231,
{\tt hep-th/9711200};
Phys. Rev. {\bf 80} (1998) 4859, {\tt hep-th/9803002} 
and references therein.

\bibitem{rey}
S.-J. Rey and J.-T. Yee, Eur. Phys. J. {\bf C22} (2001) 379,
{\tt hep-th/9803001}.


\bibitem{zahed} 
M. Rho, S.J. Sin and  I. Zahed,  Phys. Lett. {B466} (1999) 199,
{\tt hep-th/9907126} 


\bibitem{janik}
R. A. Janik and R. Peschanski,
Nucl. Phys. {\bf B565} (2000) 193;
{\tt hep-th/9907177} 

\bibitem{polchinski}
J.Polchinski and M.J. Strassler, Phys. Rev. Lett. {\bf 88} (2002)
031601, {\tt hep-th/0109174}.


\bibitem{thermo}
G.T. Horowitz and A. Strominger, { Nucl. Phys.} {\bf B360} (1991) 197.
S.S. Gubser, I.R. Klebanov and A.A. Tseytlin, {Nucl.\ Phys.\ } {\bf B534} (1998) 202;
C.P. Burgess, N.R. Constable, R.C. Myers, JHEP {\bf 1999} 9908,
C. Kim and S.J. Rey, Nucl. Phys. {\bf B564} (2000) 430.  


\bibitem{Rey_etal} 
S.-J. Rey, S. Theisen and J.-T. Yee, 
Nucl. Phys. {\bf B527} (1998) 171, {\tt hep-th/9803135}

\bibitem{PSS}
G.~Policastro, D.~T.~Son and A.~O.~Starinets,
Phys.\ Rev.\ Lett.\  {\bf 87} (2001) 081601.


\bibitem{Starinets}
A.~O.~Starinets,
Phys.\ Rev.\ D {\bf 66} (2002) 124013, {\tt hep-th/0207133}.


\bibitem{zarembo}
G.~W.~Semenoff and K.~Zarembo,
Nucl.\ Phys.\ Proc.\ Suppl.\  {\bf 108}, 106 (2002), {\tt hep-th/0202156}.

\bibitem{simonov}
Y. Simonov, {\tt hep-ph/0203059}, and references therein.

\bibitem{detar}
T.H. Hansson and I. Zahed, Nucl. Phys. {\bf B374} (1992) 277;
F.V. Koch, E.S. Shuryak, G.E Browm and A. Jackson, Phys. Rev. {\bf D46}
(1992) 3169; T.H. Hansson, M. Sporre and I. Zahed, 
Nucl. Phys. {\bf B427} (1994) 545.
\bibitem{son}
D.~T.~Son,
Phys.\ Rev.\  {\bf D59}, 094019 (1999)
[hep-ph/9812287].
 
\bibitem{strominger}
A. Strominger, Phys. Lett. {\bf B101} (1981) 271;
A.M. Polyakov, Mod. Phys. Lett. {\bf A3} (1988) 325.



\end{thebibliography}
\end{document}